\documentclass[fleqn,usenatbib]{mnras}

\usepackage{newtxtext,newtxmath}
\usepackage{ulem}

\usepackage[T1]{fontenc}

\DeclareRobustCommand{\VAN}[3]{#2}
\let\VANthebibliography\thebibliography
\def\thebibliography{\DeclareRobustCommand{\VAN}[3]{##3}\VANthebibliography}

\usepackage{grffile}

\usepackage{graphicx}	
\usepackage{amsmath}	

\title[High FIP-bias plasma]{Signature and escape of highly fractionated plasma in an active region}

\author[Brooks \& Yardley]{
David H. Brooks$^{1}$\thanks{Current address: Hinode Team, ISAS/JAXA, 3-1-1 Yoshinodai, Chuo-ku, Sagamihara, Kanagawa 252-5210, Japan}
and Stephanie L. Yardley$^{2,3}$
\\
$^{1}$College of Science, George Mason University, 4400 University Drive, Fairfax, VA 22030 USA\\
$^{2}$School of Mathematics \& Statistics, University of St Andrews, North Haugh, St Andrews, Fife, KY16 9SS\\
$^{3}$Mullard Space Science Laboratory, University College London, Holmbury St. Mary, RH5 6NT, UK
}

\date{Accepted XXX. Received YYY; in original form ZZZ}
\pubyear{2021}

\begin{document}
\label{firstpage}
\pagerange{\pageref{firstpage}--\pageref{lastpage}}
\maketitle

\begin{abstract}
Accurate forecasting of space weather requires knowledge of the source regions where solar energetic particles (SEP) and eruptive events originate.
Recent work has linked several major SEP events in 2014, January, to specific features in the host active region (AR 11944). In particular, plasma 
composition measurements in and around the footpoints of
hot, coronal loops in the core of the active region were able to explain the values later measured {\it in-situ} by the {\it Wind} spacecraft. 
Due to important differences in elemental composition between SEPs and the solar wind,
the magnitude of the Si/S elemental abundance ratio emerged as a key diagnostic of SEP seed population and solar wind source locations. We seek to understand if the results are typical
of other active regions, even if they are not solar wind sources or SEP productive. 
In this paper, we use a novel composition analysis technique, together with an evolutionary magnetic field model, in a new approach to investigate a typical solar active region (AR 11150), and identify the locations of highly fractionated (high Si/S abundance ratio) plasma.
Material confined near the footpoints of coronal loops, as in AR 11944, that in this case have expanded to the AR periphery, show the signature, and can be released from magnetic field opened by reconnection at the AR boundary.
Since the fundamental characteristics of closed field loops being opened
at the AR boundary is typical of active regions, this process is likely to be general.
\end{abstract}


\section{Introduction}
Plasma elemental composition is an important tracer of matter and particulate flow through the solar atmosphere and into the heliosphere. Relative elemental abundances show distinct patterns in the corona, SEPs, and the solar wind that can be used to understand their formation processes \citep{Meyer1985}. The fast ($>$700 km s$^{-1}$) solar wind has the same composition as the solar photosphere, but elements with a low ($<$10 eV) first ionisation potential (FIP), such as Fe, Ca, and Si, are enhanced in the corona and slow ($\sim$500 km s$^{-1}$) solar wind by factors of 2-4 compared to high FIP elements such as C, N, and O \citep{VonSteiger2000}. This is known as the FIP effect and the mechanism that preferentially transports low-FIP elements into the corona appears to operate in the chromosphere and fractionates ions from neutrals \citep{Laming2015}.

Since SEPs and the slow solar wind have a similar composition they could, in principle, emanate from the same coronal plasma source. There are two key differences, however, that suggest that SEPs originate from a separate source of pre-existing coronal material than the slow solar wind: the in-situ (SEP or solar wind) to photospheric abundance ratio dependence on FIP is not the same, and the magnitude of the abundance enhancement factor (commonly called the FIP-bias) is different for some of the same element pairs \citep{Reames2018}. 

\begin{figure*}
  \centerline{%
    \includegraphics[width=1.0\textwidth]{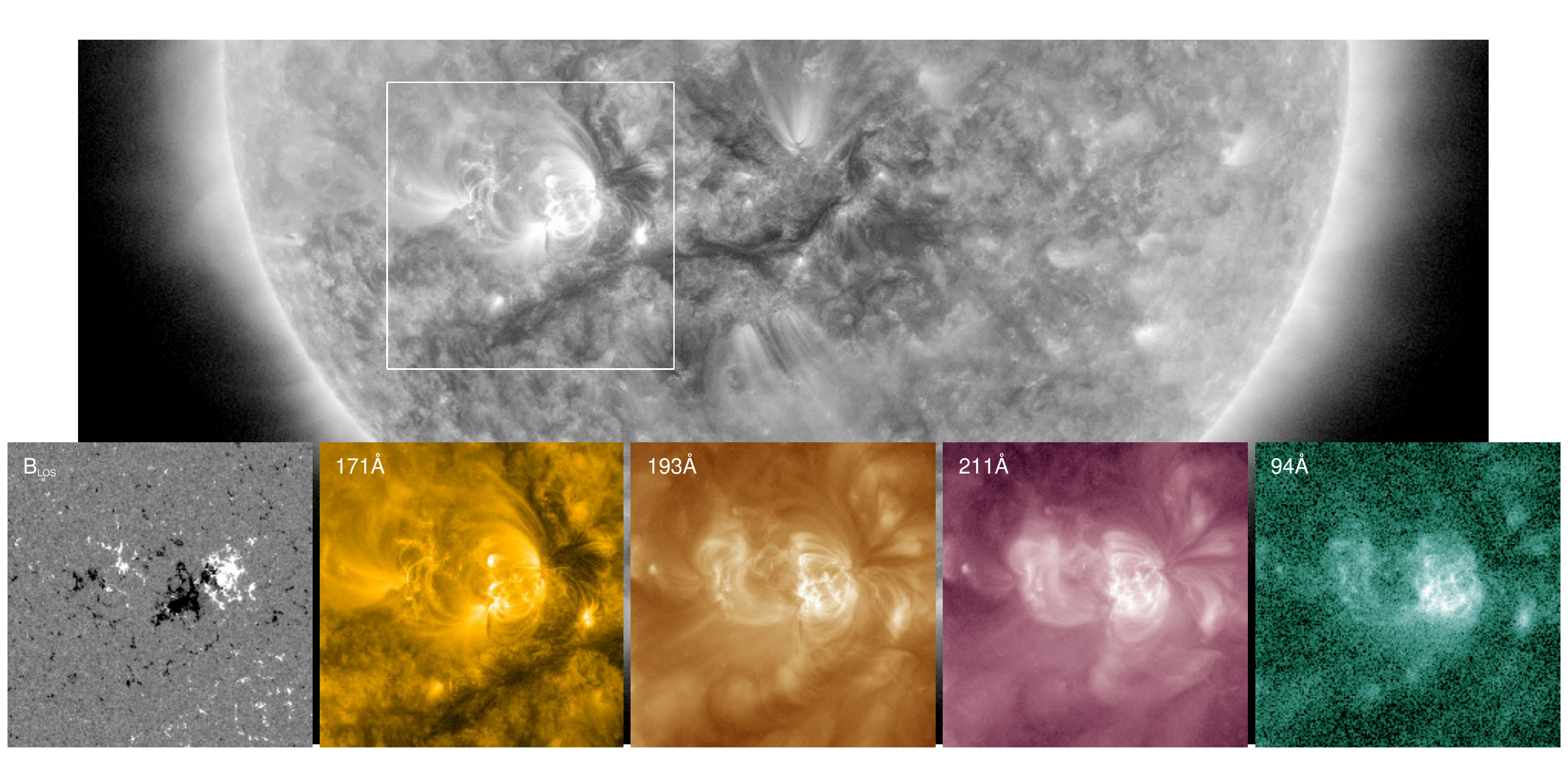}} %
  \caption{AIA images of AR 11150 observed on 2011, February 1.
The background greyscale image shows the location of the active region in the Southern hemisphere. The overlaid box shows the cutout region for the images in the bottom row. From left to right we show the longitudinal magnetic flux, 171Å, 193Å, 211Å, and 94Å images. The images are formed at temperatures in the range of 0.9—-7.1\,MK.
}
  \label{fig:fig1}
\end{figure*}

The ratio of SEP to photospheric abundance transitions from greater than one for low-FIP elements to less than one for high-FIP elements around 10 eV, whereas the transition is a few eV higher for the slow solar wind to photospheric abundance ratio \citep[see Figure 14 of ][]{Reames2018}. This means that elements close to this transition such as S (FIP = 10.36 eV) could behave differently in the two cases, providing a useful diagnostic that other elements do not. Some previous studies have argued that an abundance ratio such as Si/S shows an enhancement in SEPs, because S is behaving like a true high-FIP element, but shows no enhancement in the slow solar wind, where S behaves like a low-FIP element \citep{Reames2018}. The complete observational and theoretical picture, however, is even more complex and contradictory. The Si/S ratio does in fact often show an enhancement in the slow solar wind; albeit a little lower than detected in other element pairs. Models of the FIP effect based on the forces arising from the reflection and refraction of magnetohydrodynamic (MHD) waves \citep{Laming2004,Laming2015} show no Si/S fractionation in the slow solar wind, and a lower FIP bias than measured in SEPs. It is difficult also to reconcile these models and in-situ measurements with results from remote spectroscopic observations. The FIP bias in different coronal structures varies considerably \citep{Feldman1992}, can be larger than measured in-situ, and may also depend on the underlying technique. A recent study showed how the contributions from different components of the slow solar wind can potentially reconcile the remote and in-situ measurements \citep{Brooks2020a}, but more extensive investigations are needed.

Distilling all this confusing information down to something useful, one simple diagnostic stands out. As pointed out previously \citep{Mewaldt2002} and supported by recent studies \citep{Reames2018,Brooks2021,Harra2021}, the highest event averaged FIP-bias is seen in SEPs. More precisely, the Si/S FIP bias measured in SEPs is a factor of 1.3--2.5 higher than in the slow solar wind, where values above 2 are rarely seen in-situ for this element pair. This makes the magnitude of the FIP bias derived from Si/S lines a good diagnostic for searching for the sources of SEPs and the slow solar wind.

The EUV Imaging Spectrometer \citep[EIS,][]{Culhane2007} on Hinode \citep{Kosugi2007} observes the Si X 258.375Å and S X 264.223Å spectral lines formed in the solar corona. In combination with spectral lines from Fe ions covering a wide range of temperatures, \citep{Brooks2015} used the Si X 258.375/S X 264.223 ratio to make a map of potential slow wind sources across the entire solar disk. Following the argument that the transition from low- to high-FIP elements is higher than 10 eV in the slow wind, it was suggested that this map localises the sources of SEPs to active regions \citep{Reames2018}. A steep transition between low- and high-FIP elements has been found in AR spectra before \citep{Lanzafame2002} but a connection with SEPs was not noted, so this is a very recent idea. 

Some SEPs have been linked specifically to flares \citep{Kahler2001} and reconnection jets in sunspots \citep[][and references therein]{Nitta2006,Wang2006,Bucik2018}. 
There may also be variations that exist in the enhancement of heavy ion elements, which are potentially related to acceleration associated with magnetic reconnection during solar flares or jets. 
\citet{Mason2016}, for example, found a huge enhancement in Sulphur in 16 impulsive, $^{3}$He-rich SEP events. 
This unusual abundance signature suggests that the accelerated plasma has been heated to generate the Q/M (charge-to-mass) ratio for the C-Fe ions. Therefore, obtaining the plasma characteristics such as the temperature of the associated solar flare, can help to constrain elemental abundance measurements \citep{Bucik2021}. 
While most of the variations are systematic power laws in Q/M that are dependent upon temperature, there is also a variation that comes from the source itself.
Recently, \citet{Reames2020} has identified four distinct seed populations of SEPs however, 
when considering gradual SEPs it is likely that these events may also contain a component of reaccelerated ions 
from impulsive events or remnant suprathermal ions from previous gradual events.

In fact, only a very recent study has used EIS composition measurements to identify SEP sources \citep{Brooks2021,Brooks2021_data}. \citet{Brooks2021} used Si/S abundance ratios measured by the {\it Wind} spacecraft to trace the likely sources of several significant SEP events back to EIS Si/S composition maps of the host active region. They found that the Si/S abundance ratios measured in and around the footpoints of the hot, core loops of AR 11944 could explain the measurements made by {\it Wind}. The highest FIP bias was found closest to the region where the FIP effect operates (the top of the chromosphere) and where the plasma was confined by strong magnetic field on the order of hundreds of Gauss. The surrounding regions had a lower FIP bias, but more direct access to open field and areas of outflow, suggesting that the plasma could also form part of the solar wind.

Similar studies are difficult to perform systematically because they rely on multiple missions. EIS composition measurements have been linked
to Si/S data from the ACE observatory in the past \citep{Brooks2011}, but a hardware anomaly on ACE occurred in 2011, August, and S data are no longer
available. Due to its low abundance, S data are not routinely available from {\it Wind} either, and so we rely on large SEP events that are likely
only to occur in the active phase of the solar cycle \citep{Desai2016}. 

Nevertheless, these studies have raised questions as to whether the results are typical of ARs in general, and what the typical escape
mechanisms are for SEP and solar wind source plasma. We should also recognise the complexity of this challenge in 
times of high activity when multiple active regions are on disk. In the \citet{Brooks2021} study, one of the events was linked with a
C2.6 flare that preceeded a strong rise in the Si/S abundance ratio, but a few hours earlier a limb flare was associated with a CME from
a different region. We know that AR 11944 was producing plasma with the correct SEP Si/S composition signature, and if this material is generally 
present around the region (even if not escaping on open magnetic field), then in principle it can be accelerated by the arrival of a distant 
CME-shock; so that the host AR becomes a passive source.

This suggests other diagnostic methods we might develop to investigate further. 
Our strategy here is to use a novel EIS analysis technique that searches for higher than usual FIP bias, to see whether and where highly fractionated plasma is produced in a typical AR; regardless of whether it is SEP productive or a solar wind source. We then examine the associated magnetic field inferred from an evolutionary magnetic field model. EIS spectral lines often show high speed blue wing asymmetries \citep{Hara2008}. These are basically present in the upflows at the edges of all sizes of active regions \citep{Yardley2021b}, implying an availability of plasma with the composition 
found in the blue wing asymmetry. Our key insight is to recognize that the FIP bias in the asymmetric component will be higher if the asymmetry is larger in Si X 258.375Å than in S X 264.223Å. This is in fact what we have found in the past \citep{Brooks2012}. Therefore, it is important to look at the FIP bias as a function of velocity, since locations of higher FIP bias may be missed if the analysis is restricted to the dominant components of the spectral lines. We discuss the details of the technique, Doppler velocity calibration, and magnetic field modelling in Sect. \ref{methods}.   

In this work we present observations of AR 11150, which crossed the solar disk from 27 January to 10 February, 2011. 
In Fig. \ref{fig:fig1} we show a line-of-sight magnetogram and EUV images of the region from HMI 
\citep[Helioseismic and Magnetic Imager,][]{Schou2012} and SDO/AIA \citep[][Solar Dynamics Observaotry/Atmospheric Imaging Assembly,]{Pesnell2012,Lemen2012}. The region is magnetically simple ($\beta$-class) with leading positive polarity and trailing negative polarity sunspots connected by an arcade of coronal loops. Loops also connect to further trailing positive/negative polarities with no sunspots. Our aim is to identify the locations of highly fractionated plasma and investigate how such plasma might escape into the heliosphere, or become a reservoir that could be shock accelerated, should a CME or flare occur. We stress that according to the GOES proton flux data there are no major SEP events associated with this AR during the time-period of our observations, but we emphasize that typical AR conditions are important, since an SEP non-productive region can still become a passive source. As we approach the maximum of the new solar cycle, further investigation of the sources of significant SEP events will
be a focus of the {\it Parker Solar Probe} and {\it Solar Orbiter} missions, so another goal of this work is to test and demonstrate new analysis techniques that will support that effort.

\section{Data and Methods}
\label{methods}

\subsection{Processing details for the presented data}

We downloaded solar images from the SDO/AIA online cut-out service that accesses the Joint Science Operations Center (JSOC) at Stanford at http://jsoc.stanford.edu/. These images were processed and calibrated to level 1 using standard procedures \citep{Boerner2012}. 

The magnetograms that are taken by HMI \citep[][]{Schou2012,Couvidat2016} contain the line-of-sight component of the magnetic field and are downloaded from the JSOC website at Stanford. These magnetograms were processed and calibrated to level 1 using standard procedures although, additional pre-processing \citep[see][]{Gibb2014,Yardley2018a} is required before the magnetograms can be used as the lower boundary condition in the simulations of the coronal magnetic field. These procedures include time-averaging, the removal of low magnetic flux values and small-scale magnetic features, and flux balancing. The application of these clean-up procedures ensures that we focus on modelling the large-scale magnetic field evolution of the AR.

To account for line-of-sight effects we applied a cosine correction to the full-disk line-of-sight magnetograms to estimate the radial component of the magnetic field \citep{Yardley2018a}. We then adjust for projection effects by using a SunPy routine \citep{SunPy2020} to differentially rotate the radialised magnetograms to the time of the AR's central meridian passage. Further specific processing of the magnetograms for our modelling is described below. 

We used the Hinode/EIS for the Doppler velocity and plasma elemental composition analysis. EIS has a spectral resolution of 22mÅ and records EUV spectra in two short-wavelength (171–-212Å) and long-wavelength (245-—291Å) bands. Typically, some sub-set of these wavelength ranges containing specific spectral lines of interest is telemetered to ground. In this paper we use an EIS observing sequence that scans a field-of-view (FOV) of 240$''$ x 512$''$ using the 1$''$ slit in coarse 2$''$ steps. A 60s exposure is taken at each slit position, and about 40\% of the full spectral readout is downloaded. So a large number of diagnostic spectral lines from Fe VIII—-Fe XXIV, Ca XIV—-XVII, Si VII—-X, S X—-XIII, and Mg V—-VII are included in the observations.

A number of instrumental effects need dealt with before these data are suitable for analysis. We reduced the datasets using the standard calibration software (eis\_prep), which takes account of pixels affected by cosmic ray strikes, dust, and electrical charge.

To coalign the EIS scans with the AIA images we used the instruments' solar coordinate information to determine the approximate common FOV, then adjusted it based on image registration of bright points observed in both the AIA 193Å and EIS Fe XIII 202.044Å images. We then re-sampled the EIS FOV to the AIA plate scale and coaligned the data using cross-correlation. As a final step, we used EIS contour overlays on the original EIS data as a reference to visually adjust the EIS contour overlays on the AIA data. This helped to compensate for the fact that the EIS images are a construct of exposures taken over a 2 hour period whereas the AIA images are exposures of a few seconds.

\subsection{Relative Doppler Velocities}

Instrumental effects also affect the determination of Doppler velocities with EIS. Due to variations in the thermal conditions of the instrument as Hinode orbits the earth, there is a drift of the spectrum back and forth across the CCDs. Furthermore, the slit is not exactly aligned with the CCD axis, and also shows a small curvature. We applied an artificial neural network (ANN) model \citep{Kamio2010} to correct these effects. The ANN reproduces the orbital thermal drift by using temperature information from sensors positioned around EIS. Strong emission lines are used to accurately establish the slit tilt and curvature. After correction, the uncertainty in velocity measurements is ~4.5 km s$^{-1}$.

Since EIS lacks an absolute wavelength calibration, in this work we use Doppler velocities measured relative to a reference wavelength defined by averaging the top 50 pixels of the EIS slit across the FOV. Ideally this would be representative of quiet Sun where Doppler motions are expected to average close to zero, but in reality our observations were taken close to an active region so this is unlikely to be the case. The absolute Doppler velocities, however, are not an essential component of this work. We use the Si VII 275.368Å spectral line as wavelength standard. This is one of the strongest unblended lines in the LW spectral range where the main composition diagnostic lines we use fall.

\begin{figure}
  \centerline{%
    \includegraphics[width=0.5\textwidth]{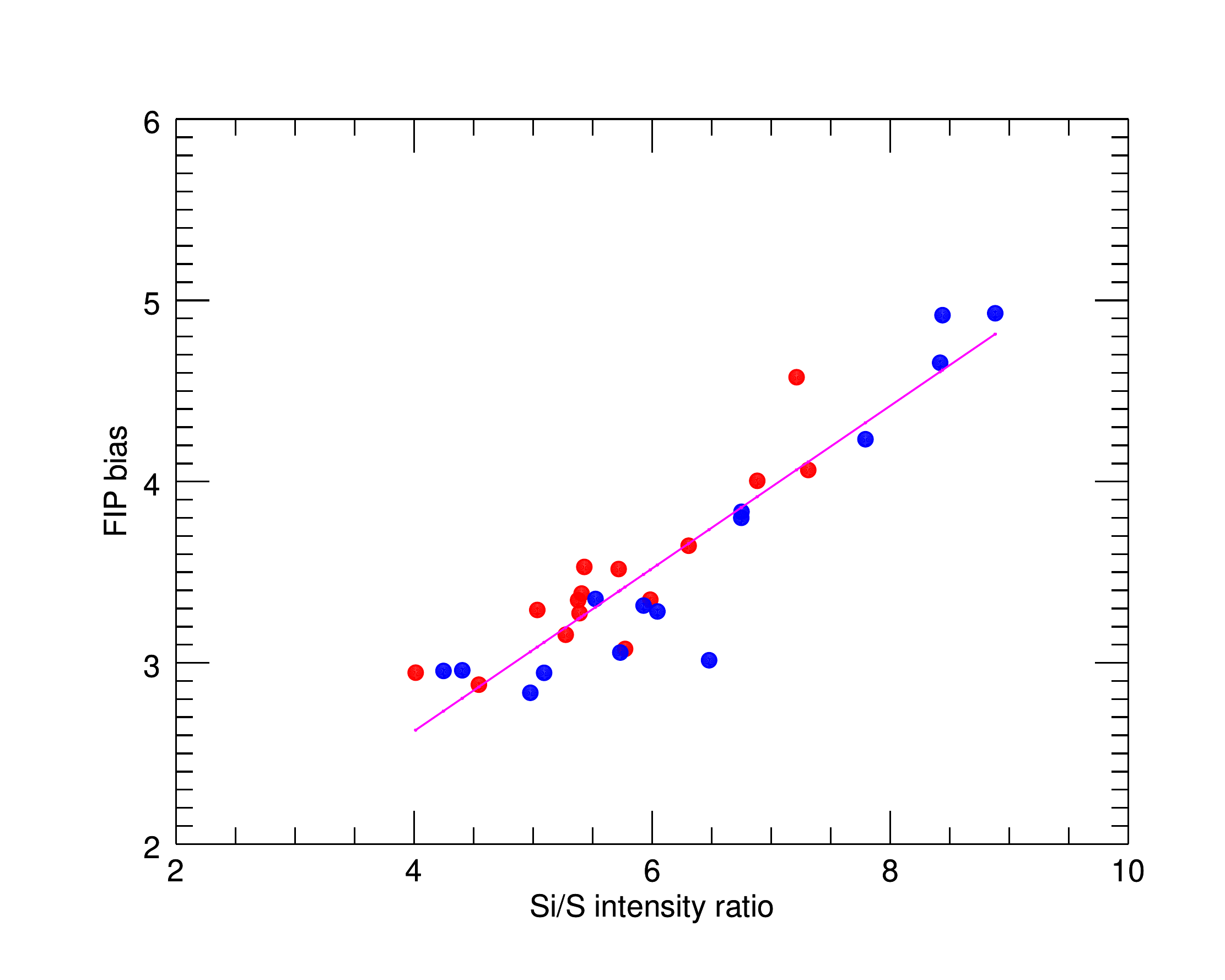}} %
  \caption{Relationship between FIP bias and Si/S intensity ratio. 
The ratio of coronal to photospheric abundance (FIP bias), measured by Hinode/EIS, as a function of the Si X 258.375Å/S X 264.233Å intensity ratio, for a sample of 30 upflow regions observed in NOAA AR 10798 in December, 2007. These upflows show high speed asymmetric components in the blue (short wavelength) wing of the line profiles. The red dots show the results for the total emission from these profiles, and the blue dots show the results for the asymmetric component only. The magenta line shows the best-fit linear relationship to the combined sample.
}
  \label{fig:fig2}
\end{figure}

\begin{figure}
  \centerline{%
    \includegraphics[width=0.5\textwidth]{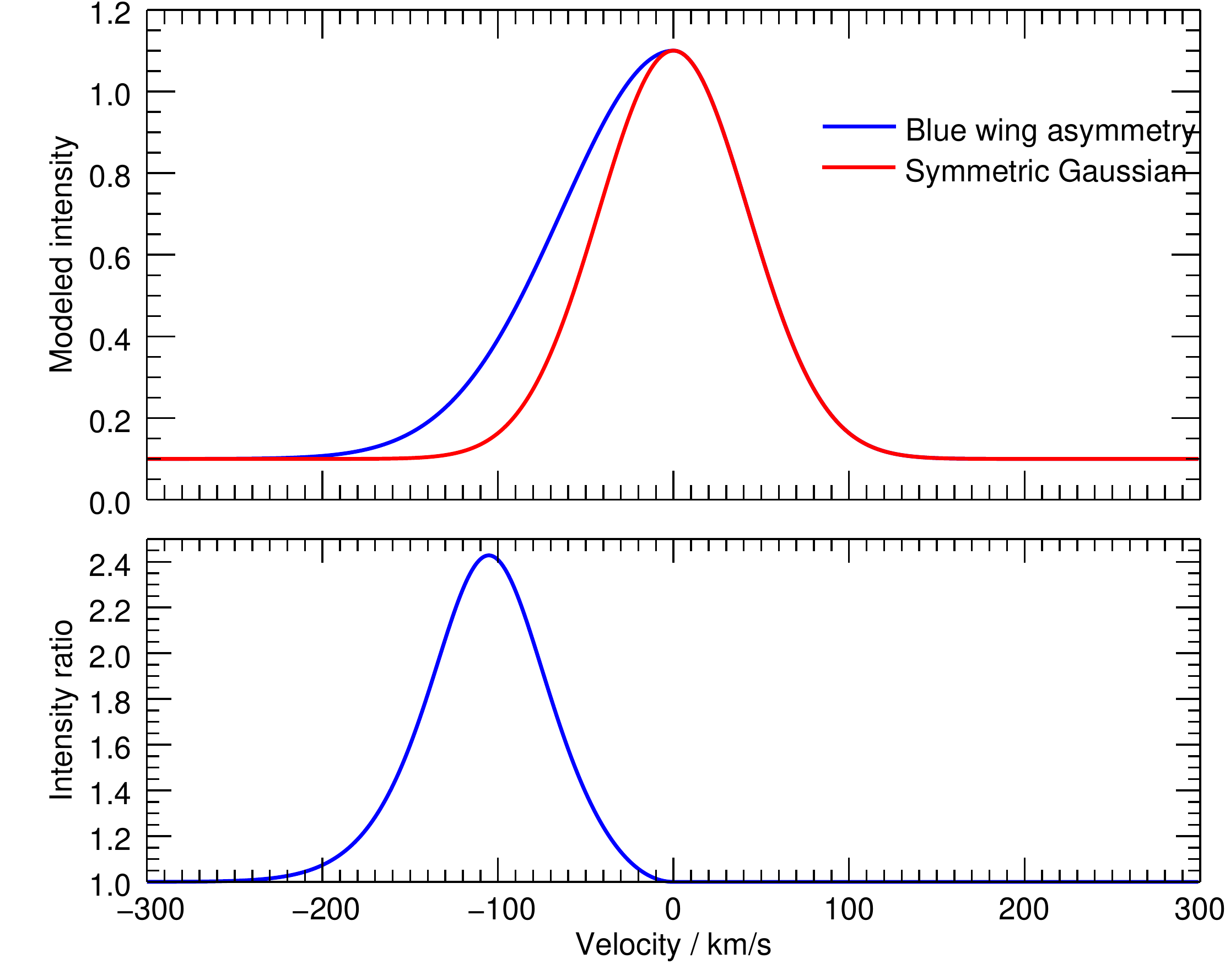}} %
  \caption{Relationship between symmetric and asymmetric Gaussian functions. The top panel shows symmetric (red) and asymmetric (blue) Gaussian functions on a velocity wavelength scale to illustrate the relationship between them. The lower panel shows the ratio of the two profiles (blue/red).
}
  \label{fig:fig3}
\end{figure}

\begin{figure}
  \centerline{%
    \includegraphics[width=0.5\textwidth]{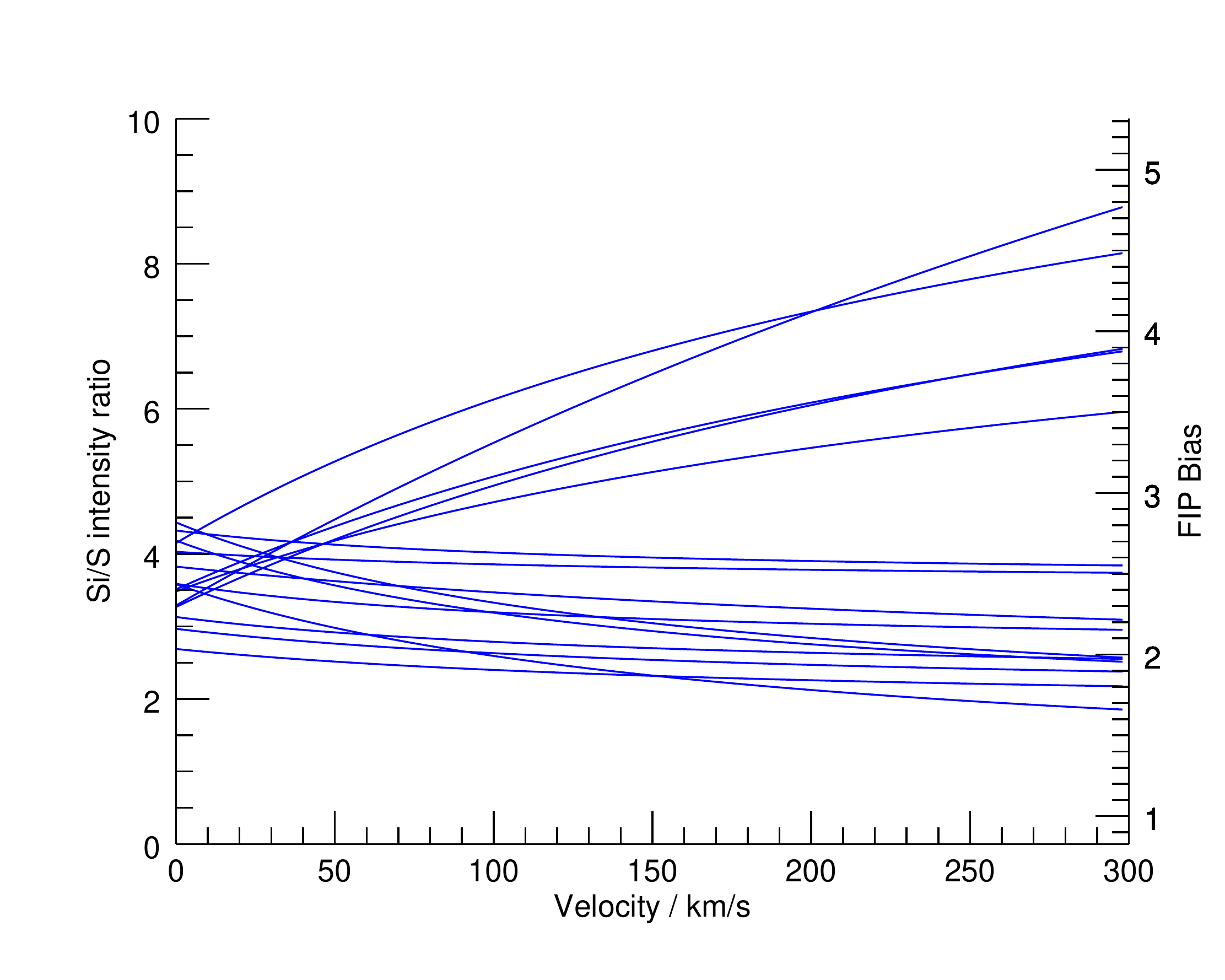}} %
  \caption{Velocity dependence of the Si/S intensity ratio.
The Si X 258.375Å/S X 264.233Å intensity ratio as a function of upflow velocity for the same upflow regions as shown in Fig. S2. We used the linear relationship from Fig. S2 to convert the ratio to FIP bias and plot the appropriate range on the right hand axis.
}
  \label{fig:fig4}
\end{figure}

\subsection{Plasma composition measurements}

The Si X 258.375Å and S X 264.223Å spectral lines are close in wavelength and formed at similar temperatures. The ratio of the two lines is therefore, in principle, a good diagnostic of the ratio of the Si and S abundance. This ratio gives a measure of the degree of fractionation of the plasma due to the FIP effect (FIP bias), though rather than being representative of the composition of all low FIP elements, in this case it properly refers to the degree of fractionation of Si compared to S. The contribution functions of the two lines also have similar temperature and density dependencies, but ideally we would derive the temperature and density structure of the observed plasma from a range of emission lines and convolve it with the contribution functions to model the ratio as accurately as possible. This is a standard analysis technique and we have used it in several previous studies of a variety of coronal features \citep{Brooks2011,Baker2015,Brooks2015}.

This method is not practical for this work. It involves integrating the intensities in wavelength across the line profiles, whereas we are interested in how the Si/S abundance ratio varies with velocity. EIS line profiles also often show asymmetries \citep{Hara2008,DePontieu2009} indicative of upflows at the 
active region edges.
Such upflows are a well known feature of many ARs and are often the locations where line profile asymmetries are most conspicuous. They appear across at least an order of magnitude in total magnetic flux \citep{Yardley2021b}, and may contribute to the slow solar wind \citep{Sakao2007,DelZanna2008,Harra2008,Doschek2008}.
In the past we have attempted to measure the plasma composition in the high-speed blue-shifted asymmetric component of the line profiles in AR upflows by fitting asymmetric and multiple Gaussian functions \citep{Brooks2012}. Two components to the line profiles may be a good approximation \citep{Tian2011}. The reality is, however, that the emission varies smoothly with wavelength, and it is not at all clear that single or multiple Gaussian fits, which necessarily have assigned velocities, will capture the variation in FIP bias over the full range of velocities apparent in the line profiles. Furthermore, the magnitude and centroid of the asymmetric component varies with temperature \citep{Brooks2012}, so it is difficult to find a solution that works for all spectral lines simultaneously.

Nevertheless, our earlier work \citep{Brooks2012} motivated this analysis, because it appears that the highest FIP bias is often detected in the asymmetric component. Fig. \ref{fig:fig2} shows the relationship between the FIP bias, derived using a full computation of the density and temperature structure, and the Si/S intensity ratio for a sample of 15 areas in the upflows associated with an AR observed in December, 2007. Details of the observations and analysis method are given in the original paper \citep{Brooks2012}. Note the almost linear relationship between the two quantities. This gives a convenient approximate method of converting between them.

To illustrate why the FIP bias is sometimes larger in the asymmetric component we show a comparison between two synthetic line profiles in Fig. \ref{fig:fig3}. The Gaussian function has a peak intensity of 1, a flat background intensity of 0.1, a full width at half maximum of 100 km s$^{-1}$, and is centred on zero velocity. The asymmetric function is formed by combining the blue and red wings of two Gaussians with different widths. The first matches the red wing of the Gaussian function, and the second has an increased width on the blue wing. The asymmetry is 20\%, which means that the difference between the widths of the Gaussian functions that fit the blue and red wings is 20\% of the total width. For further details see \citet{Brooks2012} -- especially equation 1. 

We show the ratio of the two profiles in the lower panel of Fig. \ref{fig:fig3}. As a result of the asymmetry, the intensity ratio in the blue wing shows a Gaussian distribution. In this perfect case, the intensity ratio first increases as the velocity increases, reaching a peak around 105 km s$^{-1}$, before decreasing again. Note that at all times the intensity ratio is larger than 1.

The model shows why the FIP bias is sometimes larger in the asymmetric component of the upflows. It is because in these cases the asymmetry is stronger in the Si X 258.375Å line than in the S X 264.223Å line \citep[see, for example, Fig. 2 in ][]{Brooks2012} so the intensity ratio is larger in the blue wing. In order to have photospheric abundances (FIP bias close to 1), the S X 264.223Å line needs to be relatively strong. This suggests that looking in the high velocity wings of the line profiles could be a good strategy for trying to detect high levels of FIP bias, since in some cases the asymmetry diminishes faster at longer wavelengths (larger velocities) in S X 264.223Å. The model of Fig. \ref{fig:fig3} also shows that the peak FIP bias will be associated with a specific velocity even if the asymmetry is a continuous function of velocity. This is an encouraging result for previous work that used multiple Gaussian fits. Note also that since the Si X 258.375Å and S X 264.223Å lines both show asymmetries in the blue wing, that disappear in the noise of the background spectrum at high velocities, the ratios in the real data are unlikely to match the idealised model case.

We show real examples in Fig. \ref{fig:fig4} using the same sample of AR upflows as Fig. \ref{fig:fig2}. To calibrate velocities as a function of wavelength for the Si X 258.375Å/S X 264.223Å ratio, we interpolated the line profiles to a higher resolution velocity grid (2 km s$^{-1}$) and defined the peaks of the new profiles to be zero velocity before taking the ratio of the two. We then used the linear relationship found from Fig. \ref{fig:fig2} to convert the Si/S intensity ratios to FIP bias as a function of velocity. The figure clearly shows 5 curves increasing as a function of velocity, that reach their maximum (with the highest FIP bias in the sample) at the highest velocities ($\sim$300 km s$^{-1}$). 

There are two uncertainties to note regarding this technique. First, strictly speaking the method does not necessarily find all the highest FIP bias. If there are appreciable asymmetries in the red wings of the Si X 258.375Å and S X 264.223Å lines these could produce high levels of FIP bias that would not be detected. Fortunately, detectable red wing asymmetries are not common at the formation temperatures of these lines \citep{McIntosh2012}. Second, the FIP bias is not always higher in the blue wing. In the test sample of results in Fig. \ref{fig:fig4}, the vast majority (80\%) of the curves are either flat or increasing with velocity, but a small minority (20\%) show a decrease in FIP bias. So it is most revealing to look at the actual magnitudes of the FIP bias, and for that reason we computed spatial maps of the Si/S intensity as a function of velocity for the EIS rasters shown later in Sect. \ref{results} (Fig. \ref{fig:fig6}). We use the same technique for these maps, but in this case we also filter out signals that are unlikely to be real. This was done by taking the median value of the background in the 200—-250 km s$^{-1}$ range, for each Si X 258.375Å and S X 264.223Å line profile separately, and setting the interpolated profiles to zero if the intensity is less than this value plus the uncertainty in the radiometric calibration of $\sim$22\% \citep{Lang2006}. Extreme spikes in FIP bias and values close to photospheric abundances (within the calibration uncertainty) were also filtered out.

\begin{figure*}
  \centerline{%
    \includegraphics[width=1.0\textwidth]{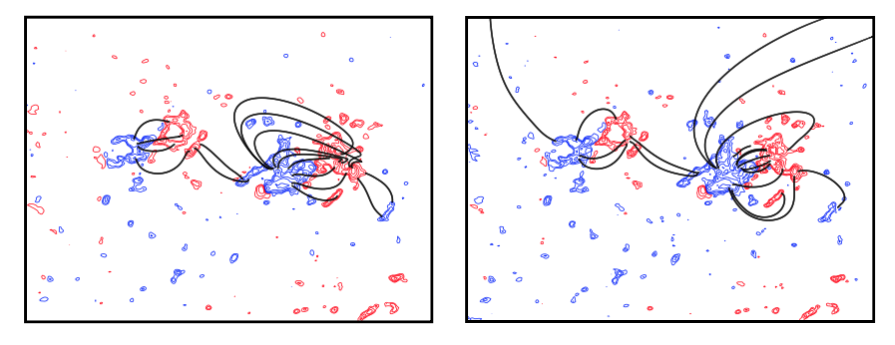}} %
  \caption{The initial condition shown for each simulation performed with: a potential field initial condition and an open top boundary (left panel), a linear force-free field initial condition and closed top boundary (right panel). Both panels show representative coronal magnetic field lines (black) taken from the simulations. The red (blue) contours represent the positive (negative) magnetic field at the photospheric surface.
}
  \label{fig:fig5}
\end{figure*}
\subsection{Evolutionary Magnetic field model}

To simulate the coronal field evolution of AR 11150 we use the time-dependent data-driven modelling technique of \citet{Mackay2011}. This technique uses a time sequence of line-of-sight magnetograms as the lower boundary condition to generate a continuous time series of non-linear force-free fields by applying magnetofrictional relaxation \citep{Yang1986}. We use the longitudinal component of the magnetic field, an initial field condition, and a time sequence of horizontal boundary motions to simulate the evolution of the coronal magnetic field. The structure of the coronal field is therefore a result of the injection of non-potentiality into the corona over hours or days due to the motions applied at the boundary. Magnetofrictional relaxation then acts to relax the coronal field, which is in non-equilibrium, into a non-linear force-free state.
Previously, this modelling technique has been tested and shown to accurately reproduce the coronal evolution and the build up of non-potential magnetic field prior to the eruption of many ARs \citep[][]{Mackay2011,Gibb2014,Yardley2018b, Yardley2021a}. More recently, the method has been used to distinguish eruptive from non-eruptive active regions for prediction purposes \citep[][Pagano et al. 2021 submitted]{Pagano2019a,Pagano2019b}.

The simulations are performed using 40 line-of-sight magnetograms taken by SDO/HMI with a cadence of 96 minutes, which have been cleaned using the clean-up procedures. The cleaned magnetograms are interpolated onto a 512$^{2}$ grid, giving a slightly lower resolution than the original line-of-sight magnetograms. The magnetograms are scaled to fill 70\% of the base of the computational box to avoid unnecessary boundary effects.

We conducted numerous simulations where we altered the initial conditions, boundary conditions and coronal parameters to find a model that best fit the observed coronal evolution of AR 11150. Here we consider two of the models that were produced, one with a linear force-free field initial condition and a closed top boundary and the other with a potential field initial condition and an open top boundary. A linear force-free field is used for the initial condition of the first simulation as the coronal loops in the observations appear to be non-potential. The initial condition is constructed from the first line-of-sight magnetogram in the sequence, which is taken on 2011, January 29, at 12:00\,UT, and is shown in Fig. \ref{fig:fig5}. The sign of the force-free parameter $\alpha$ is assigned using the sense of twist seen in the magnetic tongues in the line-of-sight magnetic field observations \citep{Luoni2011} during the AR's emergence phase. The value of $\alpha$ in the simulation scales with the size of the computational domain. In this case, we used an $\alpha$ value of -9.8$\times$10$^{-9}$ m$^{-1}$ to produce a best-fit model that closely matched the observed coronal evolution of the AR. The top and side boundaries of the computational box are closed. We also included Ohmic diffusion where the resistive coefficient $\eta$ was assigned a value of 25\,km$^{2}$ s$^{-1}$. Coronal diffusion is included to prevent the build-up of highly twisted magnetic field in the simulations as this is unrealistic. The second simulation was constructed using a potential field initial condition, an open top boundary and the inclusion of Ohmic diffusion. Due to the open nature of the top boundary the magnetograms were not flux balanced when pre-processing the magnetograms. 

\section{Results}
\label{results}

\begin{figure*}
  \centerline{%
    \includegraphics[width=1.0\textwidth]{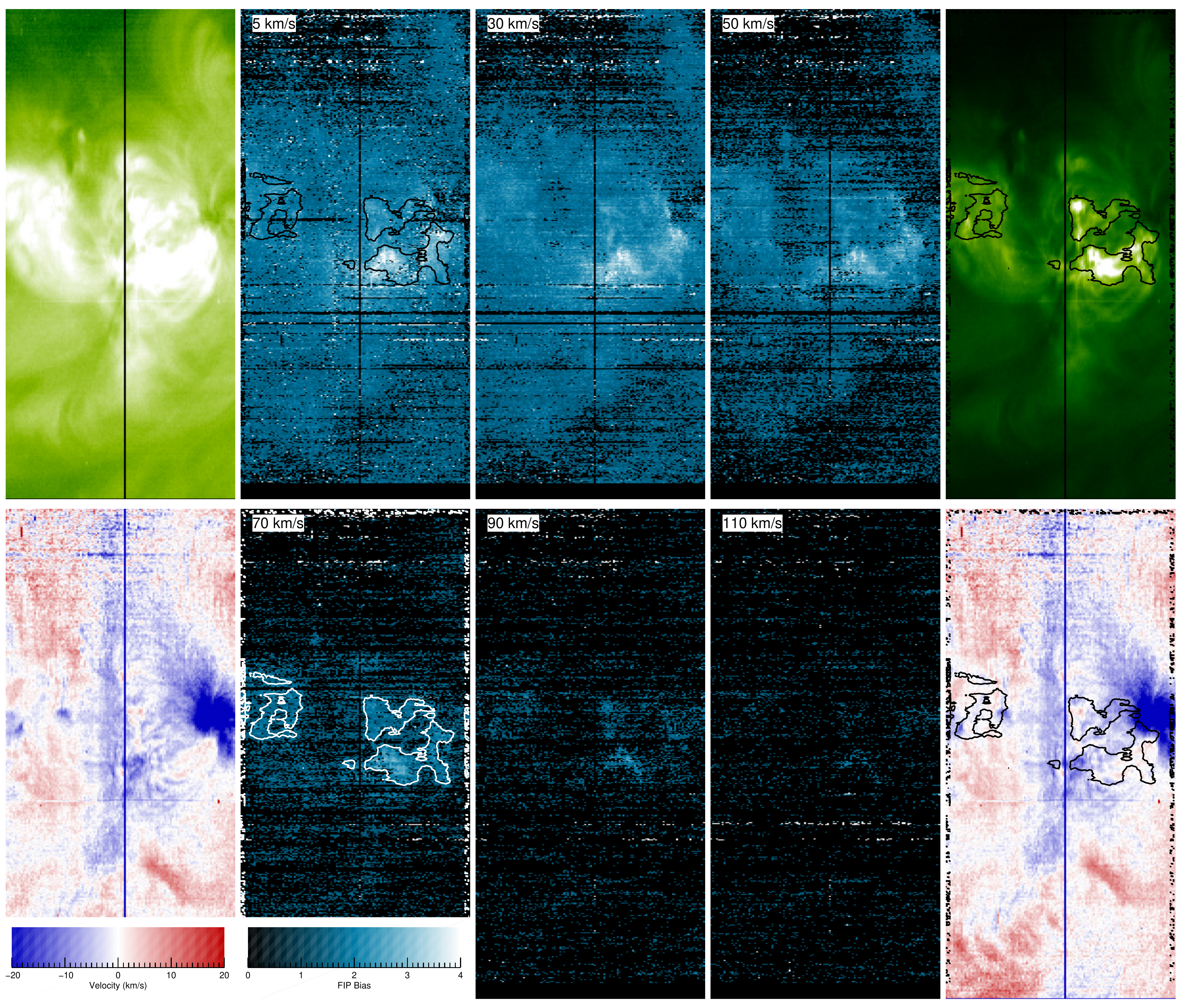}} %
  \caption{EIS observations of AR 11150 on 2011, January 31 at 10:23UT.
The left column shows an Fe XIII 202.044Å intensity image of the region (top) and a Doppler velocity map derived from the same spectral line (bottom). These are shown again in the right column with contour overlays of the brightest structures in the FIP bias maps. We show spatial FIP bias as a function of velocity in the blue images, increasing left to right and top to bottom through 5, 30, 50, 70, 90, and 110 km s$^{-1}$. We highlighted contours on the 70 km s$^{-1}$ image, and they delineate values $\sim$2.7. Colour bars show the Doppler velocity and FIP bias scales.
}
  \label{fig:fig6}
\end{figure*}

\begin{figure*}
  \centerline{%
    \includegraphics[viewport=0 80 500 425,clip]{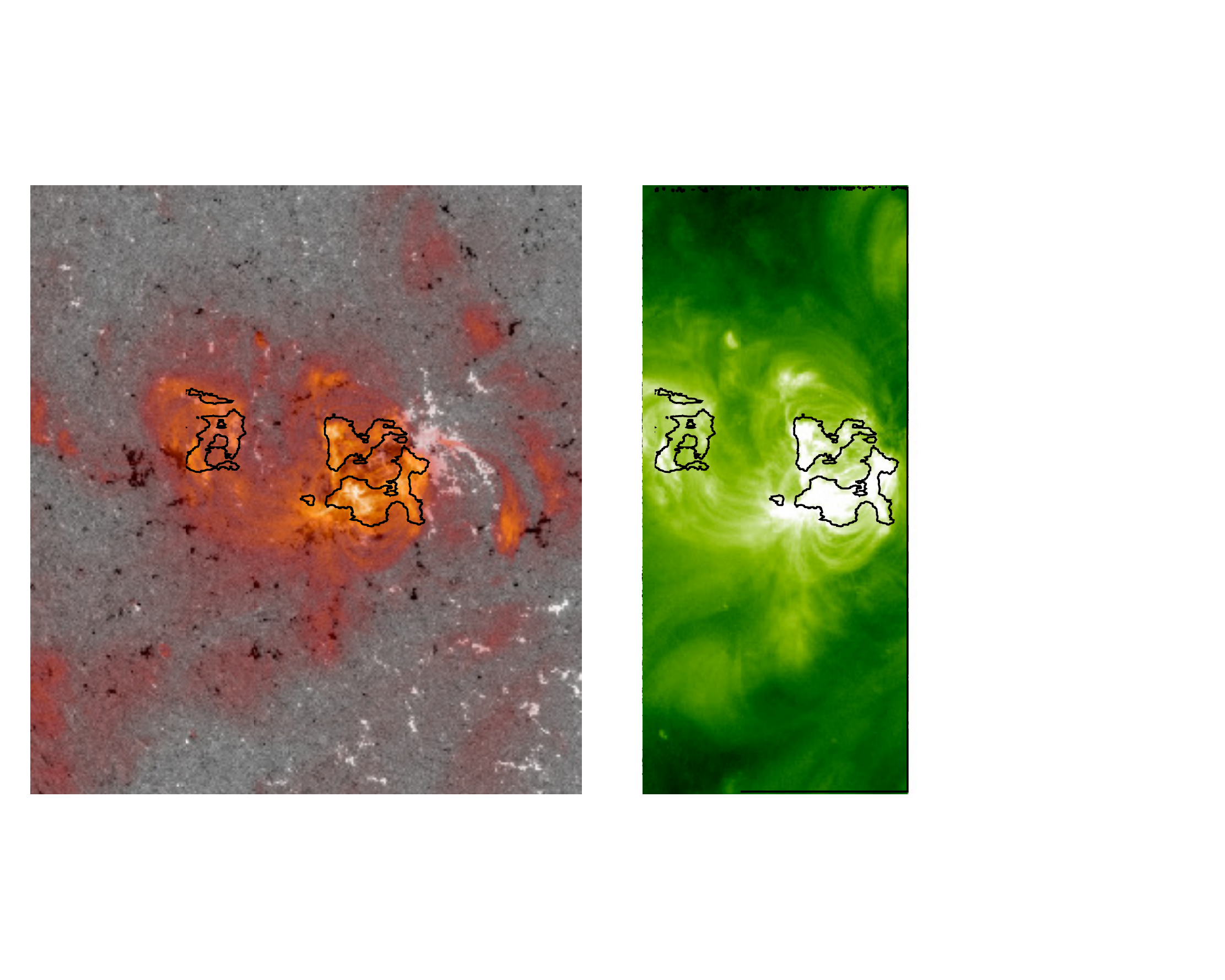}} %
  \caption{Overlays of HMI, AIA, and EIS observations of AR 11150.
The left panel shows an AIA 193Å filter image (red) coaligned and overlaid on an HMI magnetogram (black and white). The right panel shows the same AIA 193Å image (green) with the FIP bias at 70 km s$^{-1}$ from Fig. \ref{fig:fig6} overlaid as black contours.
}
  \label{fig:fig7}
\end{figure*}

\begin{figure*}
  \centerline{%
    \includegraphics[width=1.0\textwidth]{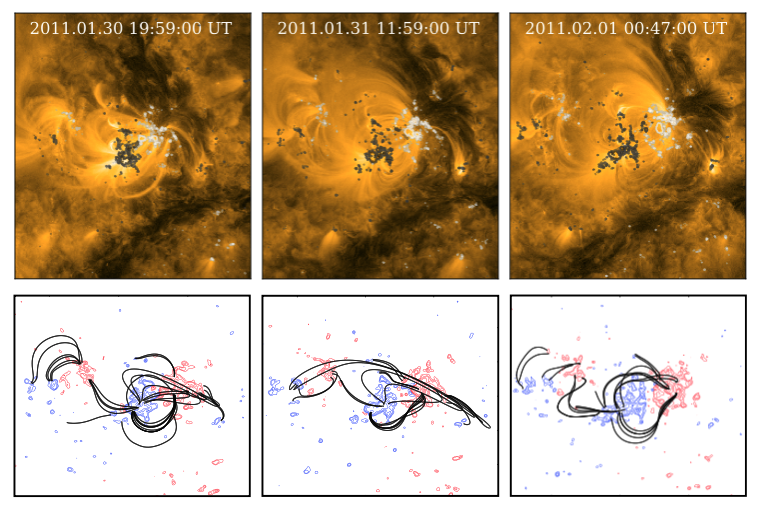}} %
  \caption{AR magnetic topology and evolution over a 29 hour period. 
The top row shows the evolution of AR 11150 in the AIA 171Å observations with the line-of-sight magnetic field from HMI overlaid on top. The white (black) contours represent the positive (negative) magnetic field of AR 11150. The bottom row shows representative magnetic field lines (black) taken from the simulation throughout the evolution. The red (blue) contours represent positive (negative) magnetic flux. The saturation of the magnetic field in both rows is set to +/- 100\,G.
}
  \label{fig:fig8}
\end{figure*}

\begin{figure*}
  \centerline{%
    \includegraphics[width=1.0\textwidth]{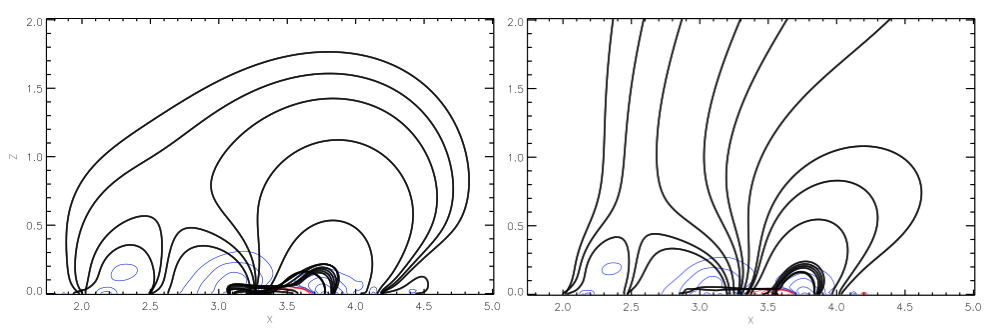}} %
  \caption{Magnetic field configuration in the xz-plane of the two magnetic field models on 31 January 11:59 UT with open (left) and closed (right) top boundary conditions. The red (blue) contours represent positive (negative) magnetic flux and the black lines represent representative field lines from the model of AR 11150.
}
  \label{fig:fig9}
\end{figure*}

\begin{figure*}
  \centerline{%
    \includegraphics[width=1.0\textwidth]{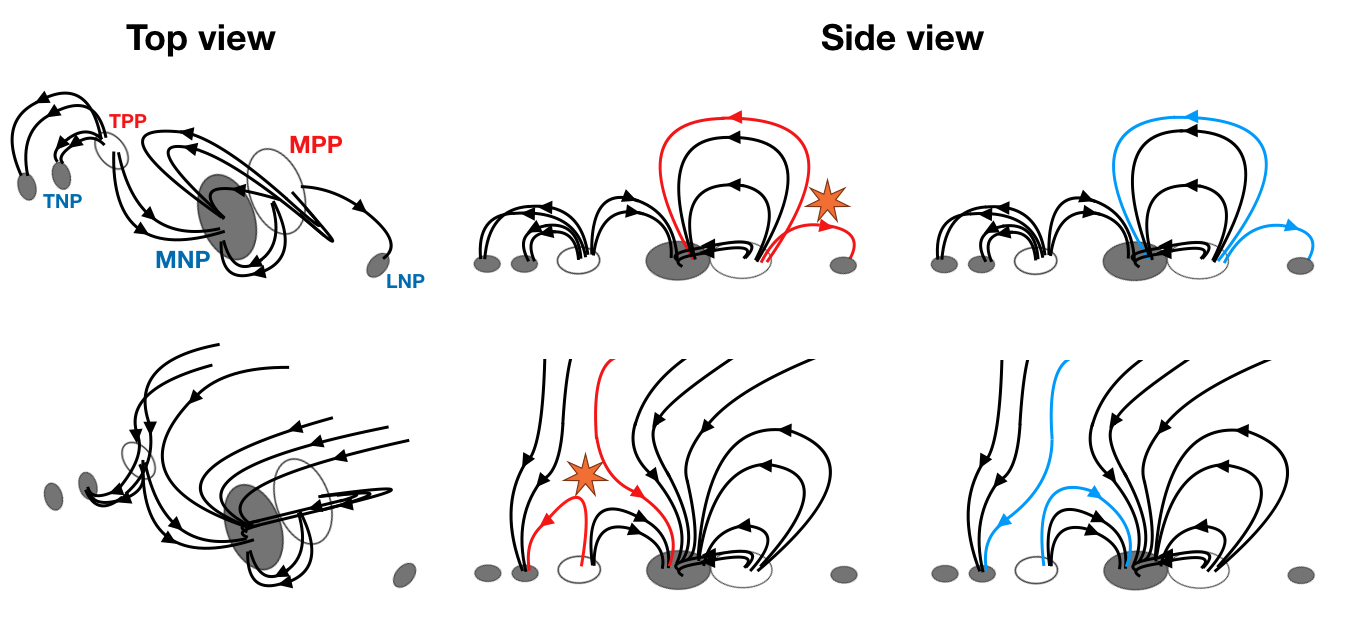}} %
  \caption{Schematic representing a top and side view of the magnetic field configuration of AR 11150 taken from the magnetic field models with closed (top panel) and open (bottom panel) top boundaries (see Fig. \ref{fig:fig9}). In the top left panel the main negative and positive polarities of the AR are denoted by MNP and MPP while the trailing negative, trailing positive and leading negative polarities are indicated by TNP, TPP and LNP, respectively. In the top panel, component reconnection (orange star) takes place between diverging closed fields located at the positive magnetic field of the active region and connecting to the negative polarity to the east and strong quiet Sun magnetic field to the west. In the bottom panel, interchange reconnection occurs between the open field associated with the negative polarity of the active region and closed magnetic field connecting with the trailing spotless positive polarity. The red lines represent the magnetic field lines involved in the component and interchange reconnection and the blue lines represent the magnetic field lines post-reconnection.
}
  \label{fig:fig10}
\end{figure*}

Here we present observations of AR 11150. The nature of our study meant that we had to be cognizant of a wide range of restrictions when selecting this AR. The observing mode and diagnostics available from EIS are generally restricted by telemetry constraints, for example. Hinode is also primarily an activity mission that targets complex and flaring ARs. There are also restrictions on the magnetic field modelling as the lower boundary conditions, which are based on the line-of-sight magnetic field measured by HMI, are affected by instrumental effects at large centre-to-limb angles. Therefore, we surveyed a sample of 20 relatively simple bipolar ARs \citep{Yardley2018a,Yardley2021a}
that emerged within $\pm$60$^o$, preferably close to disk centre, and finally settled on AR 11150. This was tracked by Hinode outside of eclipse season, and EIS observations were made with a wide field-of-view (FOV) covering most of the target region and including the important abundance diagnostic spectral lines. 

We show EIS Fe XIII 202.044Å observations in Fig \ref{fig:fig6}. This spectral line samples plasma formed around 1.8\,MK. There are strong Doppler blue-shifts (10--40 km s$^{-1}$) in regions of weak intensity at the AR boundaries to the solar west and east of the primary loop arcade connecting the two largest opposite polarities. There are also two compact blue-shifted areas near the footpoints of the loops connecting the weaker trailing opposite polarities. We have overlaid contours encapsulating the top 1/3rd of the FIP bias values on the composition and Doppler velocity maps in Fig. \ref{fig:fig6}. They show that the highest FIP bias in this region is located in bright patches close to the footpoints of the closed field loop arcades bordering the edges of the blue-shifted upflowing plasma. The high FIP-bias plasma is therefore favourably located. These patches persist in the 5-70 km s$^{-1}$ range but they weaken and their extent diminishes at higher velocities. This is consistent with our modelling of asymmetric line profile ratios (Fig. \ref{fig:fig3}).

Previous EIS observations have also shown that the highest FIP bias in a small anemone region was concentrated at the loop footpoints \citep{Baker2013}, but that was a young region emerging in a coronal hole, so the magnitude of the FIP bias was somewhat lower (less than $\sim$2). As discussed, the magnitude of the FIP bias is a useful diagnostic. Recent work has shown that high FIP bias plasma at the footpoints of the hot, core AR loops, are the likely source of several significant SEP events that occurred in 2014, January \citep{Brooks2021}, for example. Our analysis reveals that the FIP bias magnitudes in the high velocity wing of the spectral lines are also high.

In Fig. \ref{fig:fig7} we compare the locations of the highest FIP bias with an AIA 193Å image and HMI magnetogram. The footpoints of the loop arcades in the AR are rooted close to the main inversion line between the dominant polarities on the western side, and weaker scattered inversion lines on the eastern side, and in the trailing spotless magnetic concentrations. 

For further guidance on how the coronal source plasma is confined and released by the magnetic field we turn to the numerical modelling described in Sect. \ref{methods}. Recall that we used HMI line-of-sight magnetograms to simulate the evolution of the coronal magnetic field of AR 11150. We took sample magnetic field lines at different timesteps from the simulation to compare with the AIA observations of the coronal loop arcade and determine the goodness-of-fit of the model. The model that best fits the observed coronal loops in this case uses a linear force-free field initial condition, closed top boundary and includes Ohmic diffusion. We also include analysis of the model produced by using a potential field initial condition, an open top boundary and including Ohmic diffusion. 

Here we show examples of the evolution of the magnetic topology of AR 11150, from the best-fit model over a period of $\sim$30 hours surrounding the EIS observations, compared to AIA 171Å images in Fig. \ref{fig:fig8}. The representative field lines drawn from the model show a good match with the observed loop structures in the Figure. Note also the growth and expansion of the AR. The dominant magnetic polarities clearly separate between 20UT on January 30 (left panels) and 01UT on February 1 (right panels), and the loops in the arcade connecting the polarities appear longer. 

Based on theoretical models and observations, the general view is that plasma emerging with a photospheric composition needs to be confined in closed magnetic field for the FIP effect to operate. Typically it takes 2 days to reach coronal abundances, and the longer the plasma is confined the larger the FIP bias will become \citep{Widing2001}, at least until the AR begins to disperse \citep{Baker2015}. AR 11150 was visible even in far-side solar images at least a week before. Our EIS observations and modelling also suggest that as AR 11150 grows and expands, the oldest loops are pushed towards the AR periphery, so that the plasma confined longest at the footpoints, with the highest FIP bias, ends up located close to the blue-shifted upflows (Fig. \ref{fig:fig6}). The boundaries of these areas are also preferential locations for magnetic reconnection to occur that can allow the plasma to escape. This can occur in several ways and we demonstrate two possible processes here. 

In Fig. \ref{fig:fig9} we show a side-view of the modelled magnetic field configuration corresponding to the lower middle panel of Fig. \ref{fig:fig8}. Recall that this model used a closed top boundary to match the observed closed coronal loops, so there is no open field. In the right panel we show the magnetic field configuration from a second model designed to capture the open field structure of the AR. This model uses a potential field as the initial condition and has an open top boundary. In Fig. \ref{fig:fig10} we draw a schematic cartoon to highlight the two scenarios.

In the first model, the location of the diverging closed fields associated with the main positive polarity (MPP) of the AR corresponds to the location of the upflows observed to the west of the AR. Therefore, the observed upflows are also located in regions where field-opening component reconnection can occur between closed fields connecting the main positive and negative polarities of the AR (MPP and MNP), and nearly-parallel closed fields connecting to strong quiet Sun magnetic field (LNP) leading the AR to the west, as shown in Fig. \ref{fig:fig10}. In the open-field model, the location of the open field lines associated with the main negative polarity (MNP) of the AR corresponds to the upflows observed to the east of the AR, and the location of the closed field of the trailing positive polarity (TPP) corresponds to the upflows observed to its west. Therefore, interchange reconnection \citep{Crooker2002} can occur between the open field connecting to the main negative polarity (MNP) of the AR, and the oppositely-directed closed field connecting the polarities of the trailing spotless region (TNP and TPP), also shown in Fig. \ref{fig:fig10}.

\section{Summary and Discussion} \label{sec:summary}

We recently used direct comparisons of remote spectroscopic composition measurements and {\it in-situ} particle detections to identify the
likely sources of the significant SEP events from AR 11944 in 2014, January, as the footpoints of the hot, coronal loops in the core of the AR
\citep{Brooks2021}. Highly fractionated plasma was detected in these locations.
Here we have investigated AR 11150, to see whether 
similar composition signatures are seen in another typical AR, and to more fully understand the magnetic structure of the AR to see if we can
learn more about how the plasma is produced and is ultimately released.

Our observations identify specific features and locations in this AR that produce highly fractionated plasma. It appears that, in this case, the footpoints of the longest outer coronal loops of an expanding closed field arcade, contain plasma that has been confined the longest in the region closest to where the elemental fractionation mechanism operates; the top of the chromosphere -- as already discussed by \cite{Brooks2021}. The plasma confined at these footpoints therefore has the highest FIP bias. In this AR, we were able to detect it only by looking in the weaker high speed blue wing asymmetries of the EUV spectral line profiles. 

Our modelling shows multiple pathways involving interchange or component reconnection by which this plasma can eventually be released, but these processes also free confined plasma from other locations in the loop arcade (not just from the footpoints). This should lead to the footpoint material mixing with other plasma that was previously confined in the AR core. Even if the magnitude of the FIP bias is a signature of a different heliospheric component (solar wind or SEPs), and the highly fractionated plasma is formed in special conditions within the same structure, the mechanism by which this plasma escapes from the AR is the same. Our results also suggest that this is a typical
process that occurs in many active regions.

Note that we have put forward only two possible escape scenarios. The EIS Doppler map shows bulk upflows and is useful for understanding the flow structure of AR 11150 and comparison with other ARs and models in general. When looking at the FIP bias as a function of velocity, however, the contours we show in Figs. \ref{fig:fig6} and \ref{fig:fig7} indicate emission from plasma that is upflowing at 70 km s$^{-1}$. Such plasma can already outflow if the field in these regions is open by any process. It also means that there is potentially a continual supply of such material around the AR that can be accelerated as SEPs if an appropriate event occurs.

\section*{Acknowledgements}
We thank Deborah Baker for helpful comments on the manuscript. The work of D.H.B. was performed under contract to the Naval Research Laboratory and was funded by the NASA Hinode program. The work of S.L.Y. was supported by the STFC Consolidated Grant no. SMC1/YST037 and NERC via the SWIMMR Aviation Risk Modelling (SWARM) Project, grant number NE/V002899/1. Hinode is a Japanese mission developed and launched by ISAS/JAXA, collaborating with NAOJ as a domestic partner, NASA and STFC (UK) as international partners. Scientific operation of the Hinode mission is conducted by the Hinode science team organized at ISAS/JAXA. This team mainly consists of scientists from institutes in the partner countries. Support for the post-launch operation is provided by JAXA and NAOJ (Japan), STFC (U.K.), NASA, ESA, and NSC (Norway). The HMI and AIA data are courtesy of the NASA/SDO and the AIA, EVE, and HMI science teams. We would also like to thank JHelioviewer for being able to browse these data \citep[http://jhelioviewer.org,][]{Muller2017}. This research has made use of SunPy v1.0.2, an open-source and free community-developed solar data analysis Python package \citep{SunPy2020}.

\section*{Data Availability}
The EIS, AIA, and HMI data used in the paper are publically available from the instrument teams or through portals such as the Virtual Solar Observatory. The observational
and modelling data generated 
for analysis in the paper are available from the authors.

\bsp	
\label{lastpage}

\end{document}